# Algebraic solutions for $o(12) \leftrightarrow u(2) \otimes u(10)$ quantum phase transitions in the proton-neutron interacting boson model

M. M. Hammad[1,2], Andriana Martinou[1], Dennis Bonatsos[1]

[1]Institute of Nuclear and Particle Physics, National Centre for Scientific Research "Demokritos", GR- 15310 Aghia Paraskevi, Attiki, Greece
[2]Faculty of Science, Mathematics Department, Damanhour University, Egypt

**Abstract**

A simple systematic procedure to construct the proton-neutron unitary, $u_{sd}^{\pi\nu}(12)$, orthogonal, $o_{sd}^{\pi\nu}(12)$, and quasi-spin $su_{sd}^{\pi\nu}(1,1)$ algebras of the sd bosonic system is presented. New algebraic substructures of these algebras are discussed and the explicit formulae for their generators and Casimir operators are given in the spherical tensor form. The complementarity relationship of the Casimir operators of the $su_{sd}^{\pi\nu}(1,1)$ and $o_{sd}^{\pi\nu}(12)$ is derived. The exact algebraic solutions of the quantum phase transition Hamiltonian between the $o_{sd}^{\pi\nu}(12)$ and $u_s^{\pi\nu}(2) \otimes u_d^{\pi\nu}(10)$ limits has been considered, for the first time, in the framework of affine $su_{sd}^{\pi\nu}(1,1)$ Lie algebra. The low lying energy spectra of the $^{70}$Ge, $^{76-78}$Se, $^{96-98}$Mo, and $^{100-102}$Ru isotopes are calculated using the $o_{sd}^{\pi\nu}(12) \leftrightarrow u_s^{\pi\nu}(2) \otimes u_d^{\pi\nu}(10)$ transition Hamiltonian. The good agreement of our computation with empirical result in these isotopes emphasizes the importance of $u_s^{\pi\nu}(2) \otimes u_d^{\pi\nu}(10)$ limit. With this addition, symmetry can be extended to many nuclei.

## 1. Introduction

Quantum phase transitions (QPTs) have received great attention for many years in atomic nuclei [1-8]. The QPTs are the qualitative changes in the ground state properties of a system caused by variations in the coupling constants of its Hamiltonian. One typical method for exploring QPTs is to use Hamiltonians made up of two distinct sections, $H = (1 - \lambda)H_1 + \lambda H_2$. With changing the values of the control parameter $\lambda$ from 0 to 1, the spectrum of the Hamiltonian changes from the spectrum of $H_1$ to that of $H_2$.

In the algebraic model, the dynamical symmetry (DS) Hamiltonian is described in terms of Casimir operators (COs) of the algebras in a chain of the spectrum generating algebra (SGA). In this situation, the spectrum is completely solvable. The eigenfunctions and eigenenergies are labeled by a set of quantum numbers which are the irreducible representations (IRs) of the subalgebras in the chain of the SGA. The interacting boson model (IBM-1) [1, 6-8] is an algebraic model which provides a unified perspective of numerous models which existed separately. The basic assumptions of the IBM-1 are well known. The nucleus is considered as a system of $s$ ($l = 0$) and $d$ ($l = 2$) bosons which should reflect the monopole and quadrupole degrees of freedom of the underlying nucleon pairs. The IBM-1 includes the three descriptions (an anharmonic spherical vibrator, the deformed rotor-vibrator, and the γ-unstable rotor) as special cases of its Hamiltonian with DSs $u(5)$, $su(3)$, and $o(6)$, respectively. In the IBM-1, QPTs may be investigated using a Hamiltonian that interpolates between the DSs by adjusting its control parameters.

Since the beginning of the discussion of magnetic properties of collective even-even nuclei in the IBM-1, the proton-neutron version of IBM-1 had become increasingly important. The proton-neutron interacting boson model (IBM-2) [9-12] has mainly been applied to heavy nuclei where neutrons and protons are filling different orbits. Even-even nuclei were modeled in the IBM-2 as a system of proton ($\pi$) and neutron ($\nu$) bosons that interact via two-body forces. The SGA of IBM-2 is $u(12)$ as a single boson carries 12 degrees of freedom (6 from $s$ and $d$ and two from $\pi$ and $\nu$). There are the well-known $u(6) \otimes su^F(2)$ DS, with the $su^F(2)$ algebra generating $F$ spin, and the $u^\pi(6) \otimes u^\nu(6)$ DS in this model. The phase diagrams for these systems are complicated, as there are more phases and more control parameters. The phase structure of the IBM-2 is three-dimensional, represented as a tetrahedron [13, 14].

The aim of this study is twofold. The first aim is to extend the concepts of dual representations of pair of algebras [15, 16] to the two-fluid sd bosonic system. Central to our technique is the use of the Racah form of the generators. In the IBM-2, there are two important types of subalgebras for the $u_{sd}^{\pi\nu}(12)$ and $o_{sd}^{\pi\nu}(12)$. In the first type, the subalgebra is already a one-fluid two-level algebra. The second type is that in which the subalgebra is still a two-fluid but one-level algebra. A detailed comparison of the two frameworks is accomplished. Moreover, the construction of the two-fluid sd quasi-spin algebras (QSA) in two different ways is demonstrated. The relations between pairing operators and the multipole operators for the two-fluid sd bosonic system are given in a closed-form. As a result, we emphasize that the duality relations between COs of the $\mathfrak{su}_{sd}^{\pi\nu}(1,1)$ and $o_{sd}^{\pi\nu}(12)$ hold in the case of a two-fluid sd bosonic system. Some constraints on the coefficients that are used in the construction of the $o_{sd}^{\pi\nu}(12)$ are presented. We mention that for each choice of these coefficients, we have an orthogonal algebra, $o_{sd}^{\pi\nu}(12)$, these different cases being isomorphic to one another. This property does not exist only in the $o_{sd}^{\pi\nu}(12)$ but also appears in the construction of two-fluid sd-QSA.

The second aim of this study is to find exact solutions to the transition Hamiltonians between different weak and strong coupling limits of IBM-2. Exactly solvable models have been essential for understanding the physics of the quantum many-body system, particularly in circumstances where the system is highly correlated, such as atomic nuclei. In such situations, the spectrum of the Hamiltonian and the associated eigenfunctions may be determined algebraically. Examples in nuclear physics include the $\mathfrak{su}(3)$ shell model, the pure pairing model ($\mathfrak{su}(2)$ quasi-spin), and the three limits of the IBM-1. Richardson was the first to develop an exact solution to the pairing problem. This approach is today known as the Richardson-Gaudin technique [17]. Kerman quasi-spin formalism [18] was utilized to treat a large variety of cases for a limited category of seniority-conserving pairing interactions. The Bethe ansatz method [19-22] has also been used to extend the Richardson-Gaudin theory. Its significance is that the large matrix in the Fock subspace is reduced to a set of equations known as Bethe ansatz equations, the number of which equals the same number of pairs of bosons concerned. A wide range of many-body problems was recently proved to be analytically solved using infinite-dimensional algebras [23-25]. In the present study, the construction of proton-neutron infinite-dimensional algebras, $\widehat{\mathfrak{su}_{sd}^{\pi\nu}(1,1)}$, $\widehat{\mathfrak{su}_{sd}^{\pi}(1,1)}$ and $\widehat{\mathfrak{su}_{sd}^{\nu}(1,1)}$ are introduced. The exact algebraic Bethe ansatz solutions of the proton-neutron $o_{sd}^{\pi\nu}(12) \leftrightarrow \mathfrak{u}_{s}^{\pi\nu}(2) \otimes \mathfrak{u}_{d}^{\pi\nu}(10)$ transition Hamiltonians are given. This addition opens the door for understanding the properties of different phases in atomic nuclei. Using $o_{sd}^{\pi\nu}(12) \leftrightarrow \mathfrak{u}_{s}^{\pi\nu}(2) \otimes \mathfrak{u}_{d}^{\pi\nu}(10)$ transition Hamiltonian, a reasonable description of the energy levels is produced in the $^{70}$Ge, $^{76-78}$Se, $^{96-98}$Mo, and $^{100-102}$Ru isotopes.

The following sections comprise the paper. In section 2, the algebraic structure of the proton-neutron unitary and orthogonal algebras is described in detail. In section 3, the construction of the two-fluid sd-QSA in two different ways is demonstrated. The complementarity relationship of the COs of the $\mathfrak{su}_{sd}^{\pi\nu}(1,1)$ and $o_{sd}^{\pi\nu}(12)$ is derived. Section 4 is devoted to discussing the theoretical framework of the exact algebraic Bethe ansatz solutions of proton-neutron $o_{sd}^{\pi\nu}(12) \leftrightarrow \mathfrak{u}_{s}^{\pi\nu}(2) \otimes \mathfrak{u}_{d}^{\pi\nu}(10)$ transition Hamiltonian. In section 5, using $o_{sd}^{\pi\nu}(12) \leftrightarrow \mathfrak{u}_{s}^{\pi\nu}(2) \otimes \mathfrak{u}_{d}^{\pi\nu}(10)$ transition Hamiltonian, the fitting process for calculating Hamiltonian parameters and energy spectra in the $^{70}$Ge, $^{76-78}$Se, $^{96-98}$Mo, and $^{100-102}$Ru isotopes are presented. In section 6, the conclusions are briefly summarized.

## 2. The Proton-Neutron Unitary and Orthogonal Algebras

As a preliminary step toward studying the two-fluid bosonic systems, it is convenient to use spherical tensors (STs) which have well-defined properties concerning the angular momentum operators [26]. The ST $T_{\rho,l}$ of rank $l$ is defined as a set of $2(2l+1)$ components $T_{\rho,l,m}$ (where $\rho = \pi$ (proton) or $\nu$ (neutron) and $m = -l, -l+1, ..., l$) [1, 11] which satisfy the following commutation relations

$$[L_\pm, T_{\rho,l,m}] = \sqrt{l(l+1) - m(m \pm 1)} T_{\rho,l,m\pm 1}, \qquad [L_z, T_{\rho,l,m}] = m T_{\rho,l,m}, \qquad (1)$$

where $L_\pm$, and $L_z$ are spherical components of the angular momentum operators. Coupled STs $[T_{\rho_1,l_1} \otimes T_{\rho_2,l_2}]_{l_{12}}$ of two STs $T_{\rho_1,l_1}$ and $T_{\rho_2,l_2}$ is defined as the tensor of rank $l_{12}$ whose components $[T_{\rho_1,l_1} \otimes T_{\rho_2,l_2}]_{l_{12},m_{12}}$ can be expressed in terms of $T_{\rho_1,l_1,m_1}$ and $T_{\rho_2,l_2,m_2}$ according to

$$[T_{\rho_1,l_1} \otimes T_{\rho_2,l_2}]_{l_{12},m_{12}} = \sum_{m_1,m_2} C^{l_{12},m_{12}}_{l_1,m_1;l_2,m_2} T_{\rho_1,l_1,m_1} T_{\rho_2,l_2,m_2}, \qquad (2)$$

where $C^{l_{12},m_{12}}_{l_1,m_1;l_2,m_2}$ are the Clebsch Gordan coefficients (CGCs). Moreover, the coupled commutator with definite angular momentum $l_{12}$ and z component $m_{12}$ is defined as [27]

$$[T_{\rho_1,l_1}, T_{\rho_2,l_2}]_{l_{12},m_{12}} = \sum_{m_1,m_2} C^{l_{12},m_{12}}_{l_1,m_1;l_2,m_2} [T_{\rho_1,l_1,m_1}, T_{\rho_2,l_2,m_2}]. \qquad (3)$$

Using the time-reversal phase convention $\tilde{T}_{\rho_l,l,m_l} = (-1)^{l-m_l} T_{\rho_l,l,-m_l}$, the following coupled commutator factorization formula for the coupled STs is derived,

$$\left[ \left[ T^\dagger_{\rho_a,a} \otimes \tilde{T}_{\rho_b,b} \right]_e, \left[ T^\dagger_{\rho_c,c} \otimes \tilde{T}_{\rho_d,d} \right]_f \right]_g$$
$$= \left\{ (-1)^{d+g+a} \triangle_{e,f} \begin{Bmatrix} e & f & g \\ d & a & b \end{Bmatrix} \delta_{\rho_b,\rho_c} \delta_{b,c} \right\} \left[ T^\dagger_{\rho_a,a} \otimes \tilde{T}_{\rho_d,d} \right]_g$$
$$- \left\{ (-1)^{b+f+e+c} \triangle_{e,f} \begin{Bmatrix} e & f & g \\ c & b & a \end{Bmatrix} \delta_{\rho_a,\rho_d} \delta_{a,d} \right\} \left[ T^\dagger_{\rho_c,c} \otimes \tilde{T}_{\rho_b,b} \right]_g, \qquad (4)$$

where $\triangle_{ef} = \sqrt{(2e+1)(2f+1)}$, {} is $6j$ symbol and $T^\dagger_{\rho_l,l}$, and $T_{\rho_l,l}$ ($l = a, b, c, d$) are the creation and annihilation operators, respectively.

Now, let us consider the unitary algebra $\mathfrak{u}_d^{\pi\nu}(10)$, which appears in the problem of the DSs of two-fluid single-level bosonic systems, with integer angular momentum $d$ ($l = 2$). The algebra $\mathfrak{u}_d^{\pi\nu}(10)$ is spanned by the generators

$$G^e_{\pi\pi,dd} = [d^\dagger_\pi \otimes \tilde{d}_\pi]_e \quad (e = 0,1,...,4), \tag{5.1}$$

$$G^e_{\nu\nu,dd} = [d^\dagger_\nu \otimes \tilde{d}_\nu]_e \quad (e = 0,1,...,4), \tag{5.2}$$

$$G^e_{\pi\nu,dd} = [d^\dagger_\pi \otimes \tilde{d}_\nu]_e \quad (e = 0,1,...,4), \tag{5.3}$$

$$G^e_{\nu\pi,dd} = [d^\dagger_\nu \otimes \tilde{d}_\pi]_e \quad (e = 0,1,...,4). \tag{5.4}$$

From the complete set of STs in (5), we can show that the 25 STs (5.1) and (5.2) can be regarded as the generators of the one-fluid unitary algebras $\mathfrak{u}_d^\pi(5)$ and $\mathfrak{u}_d^\nu(5)$, respectively. The coupled commutators of these generators with $(\rho, \acute{\rho}) = (\pi, \pi)$ or $(\nu, \nu)$ are

$$\left[[d^\dagger_\rho \otimes \tilde{d}_{\acute{\rho}}]_e, [d^\dagger_\rho \otimes \tilde{d}_{\acute{\rho}}]_f\right]_g = \{1 - (-1)^{f+e-g}\}(-1)^g \triangle_{ef} \begin{Bmatrix} e & f & g \\ 2 & 2 & 2 \end{Bmatrix} [d^\dagger_\rho \otimes \tilde{d}_{\acute{\rho}}]_g, \tag{6}$$

The STs from different subalgebras, $\mathfrak{u}_d^\pi(5)$ and $\mathfrak{u}_d^\nu(5)$, commute and together form the generators for the direct product $\mathfrak{u}_d^\pi(5) \otimes \mathfrak{u}_d^\nu(5)$ algebra. If $e$ is an odd number, the generators (5.1), and (5.2) span the one-fluid orthogonal algebras $o_d^\pi(5)$ and $o_d^\nu(5)$, respectively. The generators of $o_d^\pi(5)$ and $o_d^\nu(5)$ commute and together form the algebra $o_d^\pi(5) \otimes o_d^\nu(5)$.

This result can be extended to the $\mathfrak{u}_{sd}^{\pi\nu}(12)$ algebra which appears in the problem of the DSs of two-fluid two-level systems, with integer angular momenta $s$, and $d$. Specifically, the algebra $\mathfrak{u}_{sd}^{\pi\nu}(12)$ is spanned by the generators

$$G^0_{\rho\acute{\rho},ss} = [s^\dagger_\rho \otimes \tilde{s}_{\acute{\rho}}]_0, \tag{7.1}$$

$$G^{0,1,...,4}_{\rho\acute{\rho},dd} = [d^\dagger_\rho \otimes \tilde{d}_{\acute{\rho}}]_{0,1,...,4}, \tag{7.2}$$

$$G^2_{\rho\acute{\rho},sd} = [s^\dagger_\rho \otimes \tilde{d}_{\acute{\rho}}]_2, \tag{7.3}$$

$$G^2_{\rho\acute{\rho},ds} = [d^\dagger_\rho \otimes \tilde{s}_{\acute{\rho}}]_2, \tag{7.4}$$

where $(\rho, \acute{\rho}) = \{(\pi, \pi), (\nu, \nu), (\pi, \nu), (\nu, \pi)\}$.

There are two types of subalgebras of particular importance. In the first type, the subalgebra is still a two-fluid Lie algebra formed by (7.1)-(7.4). The generators (7.1) and (7.2) constitute the two-fluid single-level algebras $\mathfrak{u}_s^{\pi\nu}(2)$, and $\mathfrak{u}_d^{\pi\nu}(10)$, respectively. The STs from different subalgebras $\mathfrak{u}_s^{\pi\nu}(2)$ and $\mathfrak{u}_d^{\pi\nu}(10)$ commute, and together form the generators for the direct product $\mathfrak{u}_s^{\pi\nu}(2) \otimes \mathfrak{u}_d^{\pi\nu}(10)$ algebra. The second type is that in which the subalgebra is one-fluid Lie algebra. The generators (7.1)-(7.4) define the boson representation of the one-fluid two-levels algebras $\mathfrak{u}_{sd}^\pi(6)$ with $(\rho, \acute{\rho}) = (\pi, \pi)$ and $\mathfrak{u}_{sd}^\nu(6)$ with $(\rho, \acute{\rho}) = (\nu, \nu)$. The STs from subalgebras $\mathfrak{u}_{sd}^\pi(6)$ and $\mathfrak{u}_{sd}^\nu(6)$ commute, and together form the generators for the direct product $\mathfrak{u}_{sd}^\pi(6) \otimes \mathfrak{u}_{sd}^\nu(6)$ algebra. Consequently, depending on the method of reconstruction of the subalgebras, there exist different coupling schemes to construct subalgebras from the generators of the two-fluid algebra $\mathfrak{u}_{sd}^{\pi\nu}(12)$. We pay attention to the following schemes

$$\mathfrak{u}_s^\pi(1) \otimes \mathfrak{u}_s^\nu(1) \subset \mathfrak{u}_s^{\pi\nu}(2), \quad \mathfrak{u}_d^\pi(5) \otimes \mathfrak{u}_d^\nu(5) \subset \mathfrak{u}_d^{\pi\nu}(10), \quad \mathfrak{u}_s^{\pi\nu}(2) \otimes \mathfrak{u}_d^{\pi\nu}(10) \subset \mathfrak{u}_{sd}^{\pi\nu}(12), \tag{8.1}$$

$$\mathfrak{u}_s^\pi(1) \otimes \mathfrak{u}_d^\pi(5) \subset \mathfrak{u}_{sd}^\pi(6), \quad \mathfrak{u}_s^\nu(1) \otimes \mathfrak{u}_d^\nu(5) \subset \mathfrak{u}_{sd}^\nu(6), \quad \mathfrak{u}_{sd}^\pi(6) \otimes \mathfrak{u}_{sd}^\nu(6) \subset \mathfrak{u}_{sd}^{\pi\nu}(12). \tag{8.2}$$

Accordingly, we have the following subalgebra chains

$$\mathfrak{u}_{sd}^{\pi\nu}(12) \supset \begin{Bmatrix} \mathfrak{u}_d^{\pi\nu}(10) \otimes \mathfrak{u}_s^{\pi\nu}(2) \\ \mathfrak{u}_{sd}^\pi(6) \otimes \mathfrak{u}_{sd}^\nu(6) \end{Bmatrix} \supset \mathfrak{u}_d^\pi(5) \otimes \mathfrak{u}_d^\nu(5). \tag{9}$$

This result explains that there are two types of weak-coupling DS limits. This algebraic structure is portrayed schematically in Figure 1.

To write the quadratic CO of $\mathfrak{u}_{sd}^{\pi\nu}(12)$ in more physical form, the following operators are introduced

$$Q_{ss}^{\pi\nu} = 2N_s^\pi N_s^\nu + N_s^\pi + N_s^\nu, \tag{10.1}$$

$$Q_{dd}^{\pi\nu} = 2N_d^\pi N_d^\nu + 5N_d^\pi + 5N_d^\nu, \tag{10.2}$$

$$Q_{sd}^{\pi\pi} = 2N_s^\pi N_d^\pi + 5N_s^\pi + N_d^\pi, \tag{10.3}$$

$$Q_{sd}^{\nu\nu} = 2N_s^\nu N_d^\nu + 5N_s^\nu + N_d^\nu, \tag{10.4}$$

$$Q_{sd}^{\pi\nu} = 2N_s^\pi N_d^\nu + 5N_s^\pi + N_d^\nu, \tag{10.5}$$

$$Q_{sd}^{\nu\pi} = 2N_s^\nu N_d^\pi + 5N_s^\pi + N_d^\pi, \tag{10.6}$$

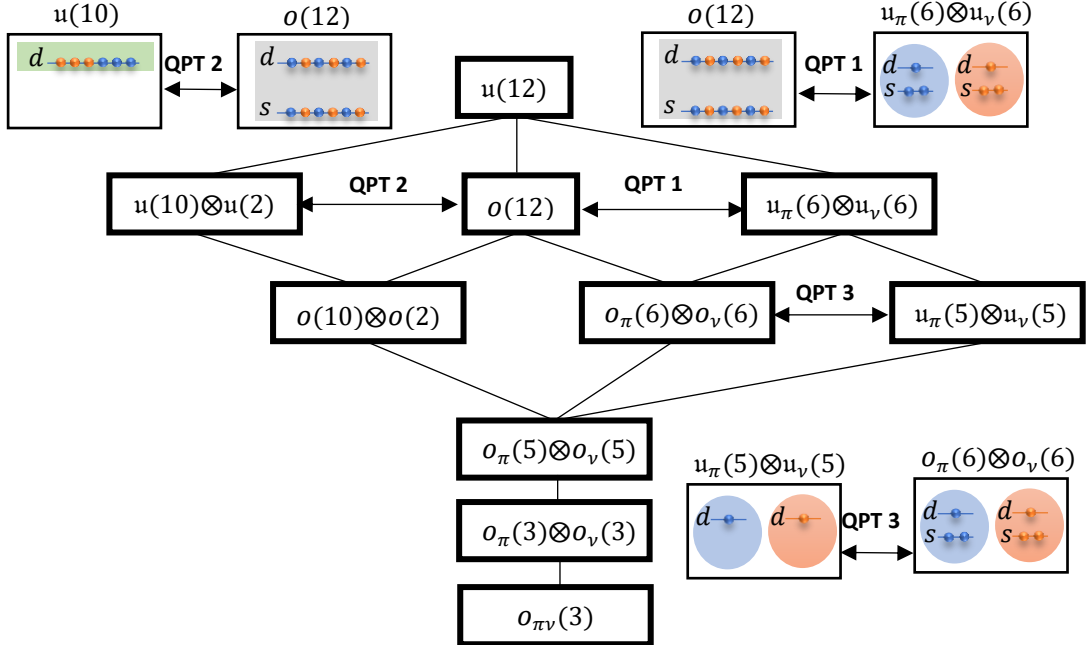

Figure 1. Different DSs and QPTs between weak and strong coupling limits of the IBM-2.

where $N_s^\pi$, $N_s^\nu$, $N_d^\pi$ and $N_d^\nu$ are the one-fluid single-level number operators. The operators $Q$ are related to the multipole operators, $\bar{Q}$, of the $o_{sd}^{\pi\nu}(12)$ as we will see later (21). So, we can call these operators "unitary multipole operators". Depending on the method of reconstruction of the subalgebras, a brief calculation shows that

$$C_2[\mathfrak{u}_{sd}^{\pi\nu}(12)] = C_2[\mathfrak{u}_d^{\pi\nu}(10)] + C_2[\mathfrak{u}_s^{\pi\nu}(2)] + Q_{sd}^{\pi\pi} + Q_{sd}^{\nu\nu} + Q_{sd}^{\pi\nu} + Q_{sd}^{\nu\pi}, \quad (11.1)$$

$$C_2[\mathfrak{u}_{sd}^{\pi\nu}(12)] = C_2[\mathfrak{u}_{sd}^{\pi}(6)] + C_2[\mathfrak{u}_{sd}^{\nu}(6)] + Q_{ss}^{\pi\nu} + Q_{dd}^{\pi\nu} + Q_{sd}^{\pi\nu} + Q_{sd}^{\nu\pi}. \quad (11.2)$$

Proceeding to the orthogonal algebra, let us consider a two-fluid single-level orthogonal algebra. The 45 generators

$$G_{\pi\pi,dd}^e = [d_\pi^\dagger \otimes \tilde{d}_\pi]_e \qquad (e = 1,3), \quad (12.1)$$

$$G_{\nu\nu,dd}^e = [d_\nu^\dagger \otimes \tilde{d}_\nu]_e \qquad (e = 1,3), \quad (12.2)$$

$$F_{\pi\nu,dd}^e = \eta_{\pi\nu,dd}^e [d_\pi^\dagger \otimes \tilde{d}_\nu]_e + \xi_{\pi\nu,dd}^e [d_\nu^\dagger \otimes \tilde{d}_\pi]_e \qquad (e = 0,1,\dots,4), \quad (12.3)$$

form a subalgebra of the $\mathfrak{u}_d^{\pi\nu}(10)$ which is referred to as the $o_d^{\pi\nu}(10)$. The coefficients of (12.3) can be determined by the requirement of closure of the commutation relation. The closure is obtained if the coefficients $\eta_{\pi\nu,dd}^e$ and $\xi_{\pi\nu,dd}^e$ satisfy the following conditions $\frac{\xi_{\pi\nu,dd}^e}{\eta_{\pi\nu,dd}^e} = (-1)^e \sigma_{dd}^{\pi\nu}$, $\sigma_{dd}^{\pi\nu} = \pm 1$. The operators $G_{\pi\pi,dd}^e \circ G_{\pi\pi,dd}^e$ and $G_{\nu\nu,dd}^e \circ G_{\nu\nu,dd}^e$ defined by

$$G_{\pi\pi,dd}^e \circ G_{\pi\pi,dd}^e = -\sum_{e\ odd} \Delta_e \left[[d_\pi^\dagger \otimes \tilde{d}_\pi]_e \otimes [d_\pi^\dagger \otimes \tilde{d}_\pi]_e\right]_0, \quad (13.1)$$

$$G_{\nu\nu,dd}^e \circ G_{\nu\nu,dd}^e = -\sum_{e\ odd} \Delta_e \left[[d_\nu^\dagger \otimes \tilde{d}_\nu]_e \otimes [d_\nu^\dagger \otimes \tilde{d}_\nu]_e\right]_0, \quad (13.2)$$

commutes with all the generators of $o_d^{\pi\nu}(10)$. The normalized quadratic CO of $o_d^{\pi\nu}(10)$ is given by

$$C_2[o_d^{\pi\nu}(10)] = 4G_{\pi\pi,dd}^e \circ G_{\pi\pi,dd}^e + 4G_{\nu\nu,dd}^e \circ G_{\nu\nu,dd}^e + 2F_{\pi\nu,dd}^e \circ F_{\pi\nu,dd}^e. \quad (14)$$

This algebraic structure can be extended to two-fluid sd bosonic system. The $o_{sd}^{\pi\nu}(12)$ is spanned by the generators

$$F_{\pi\nu,ss}^e = \eta_{\pi\nu,ss}^e [s_\pi^\dagger \otimes \tilde{s}_\nu]_e + \xi_{\pi\nu,ss}^e [s_\nu^\dagger \otimes \tilde{s}_\pi]_e \qquad (e = 0), \quad (15.1)$$

$$G_{\pi\pi,dd}^e = [d_\pi^\dagger \otimes \tilde{d}_\pi]_e \qquad (e = 1,3), \quad (15.2)$$

$$G_{\nu\nu,dd}^e = [d_\nu^\dagger \otimes \tilde{d}_\nu]_e \qquad (e = 1,3), \quad (15.3)$$

$$F_{\pi\nu,dd}^e = \eta_{\pi\nu,dd}^e [d_\pi^\dagger \otimes \tilde{d}_\nu]_e + \xi_{\pi\nu,dd}^e [d_\nu^\dagger \otimes \tilde{d}_\pi]_e \qquad (e = 0,1,\dots,4), \quad (15.4)$$

$$F_{\pi\pi,sd}^e = \eta_{\pi\pi,sd}^e [s_\pi^\dagger \otimes \tilde{d}_\pi]_e + \xi_{\pi\pi,sd}^e [d_\pi^\dagger \otimes \tilde{s}_\pi]_e \qquad (e = 2), \quad (15.5)$$

$$F_{\nu\nu,sd}^e = \eta_{\nu\nu,sd}^e [s_\nu^\dagger \otimes \tilde{d}_\nu]_e + \xi_{\nu\nu,sd}^e [d_\nu^\dagger \otimes \tilde{s}_\nu]_e \qquad (e = 2), \quad (15.6)$$

$$F_{\pi\nu,sd}^e = \eta_{\pi\nu,sd}^e [s_\pi^\dagger \otimes \tilde{d}_\nu]_e + \xi_{\pi\nu,sd}^e [d_\nu^\dagger \otimes \tilde{s}_\pi]_e \qquad (e = 2), \quad (15.7)$$

$$F_{\nu\pi,sd}^e = \eta_{\nu\pi,sd}^e [s_\nu^\dagger \otimes \tilde{d}_\pi]_e + \xi_{\nu\pi,sd}^e [d_\pi^\dagger \otimes \tilde{s}_\nu]_e \qquad (e = 2), \quad (15.8)$$

and the coefficients of (15.1), (15.4)-(15.8) can be determined by the requirement of closure of the commutation relation. The closure is obtained if the coefficients $\eta$ and $\xi$ satisfy the following conditions

$$\frac{\xi^e_{\pi v,ss}}{\eta^e_{\pi v,ss}} = (-1)^e \sigma^{\pi v}_{ss}, \frac{\xi^e_{\pi v,dd}}{\eta^e_{\pi v,dd}} = (-1)^e \sigma^{\pi v}_{dd}, \frac{\xi^e_{\pi\pi,sd}}{\eta^e_{\pi\pi,sd}} = (-1)^e \sigma^{\pi\pi}_{sd},$$

$$\frac{\xi^e_{vv,sd}}{\eta^e_{vv,sd}} = (-1)^e \sigma^{vv}_{sd}, \frac{\xi^e_{\pi v,sd}}{\eta^e_{\pi v,sd}} = (-1)^e \sigma^{\pi v}_{sd}, \frac{\xi^e_{v\pi,sd}}{\eta^e_{v\pi,sd}} = (-1)^e \sigma^{v\pi}_{sd}, \tag{16.1}$$

$$\sigma^{\pi v}_{ss}\sigma^{v\pi}_{sd} = -\sigma^{\pi\pi}_{sd}, \quad \sigma^{\pi v}_{ss}\sigma^{vv}_{sd} = -\sigma^{\pi v}_{sd}, \quad \sigma^{\pi\pi}_{sd}\sigma^{\pi v}_{dd} = -\sigma^{\pi v}_{sd}, \quad \sigma^{v\pi}_{sd}\sigma^{\pi v}_{dd} = -\sigma^{vv}_{sd}. \tag{16.2}$$

Before proceeding to analyze the subalgebras structure of the two-fluid algebra $o^{\pi v}_{sd}(12)$, we mention that for each choice of the set of coefficients $\eta$ and $\sigma$, we have a two-fluid algebra $o^{\pi v}_{sd}(12)$, these different cases are isomorphic to one another. Table 1 shows the possible values of the coefficients $\sigma$ when we combine the conditions (16.2). There are eight methods to generate two-fluid algebra $o^{\pi v}_{sd}(12)$. This property does not exist only in $o^{\pi v}_{sd}(12)$ but also appears in the construction of two-fluid quasi-spin algebra. Hence, for each $o^{\pi v}_{sd}(12)$ there is a corresponding two-fluid quasi-spin algebra. In the remaining of this paper, we choose the coefficients $\sigma$ and $\eta$ to be

$$\eta^e_{\pi v,ss} = \eta^e_{\pi v,dd} = \eta^e_{\pi\pi,sd} = \eta^e_{vv,sd} = \eta^e_{\pi v,sd} = \eta^e_{v\pi,sd} = 1, \tag{17.1}$$

$$\sigma^{\pi v}_{ss} = \sigma^{\pi v}_{dd} = \sigma^{\pi\pi}_{sd} = \sigma^{vv}_{sd} = \sigma^{\pi v}_{sd} = \sigma^{v\pi}_{sd} = -1. \tag{17.2}$$

As a result, the $o^{\pi v}_{sd}(12)$ is spanned by the two sets of single-level generators (15.2), and (15.3), and the following sets of mixed generators

$$F^e_{\pi v,ss} = [s^\dagger_\pi \otimes \tilde{s}_v]_e - (-1)^e [s^\dagger_v \otimes \tilde{s}_\pi]_e \qquad (e=0), \tag{18.1}$$

$$F^e_{\pi v,dd} = [d^\dagger_\pi \otimes \tilde{d}_v]_e - (-1)^e [d^\dagger_v \otimes \tilde{d}_\pi]_e \qquad (e=0,1,\ldots,4), \tag{18.2}$$

$$F^e_{\pi\pi,sd} = [s^\dagger_\pi \otimes \tilde{d}_\pi]_e - (-1)^e [d^\dagger_\pi \otimes \tilde{s}_\pi]_e \qquad (e=2), \tag{18.3}$$

$$F^e_{vv,sd} = [s^\dagger_v \otimes \tilde{d}_v]_e - (-1)^e [d^\dagger_v \otimes \tilde{s}_v]_e \qquad (e=2), \tag{18.4}$$

$$F^e_{\pi v,sd} = [s^\dagger_\pi \otimes \tilde{d}_v]_e - (-1)^e [d^\dagger_v \otimes \tilde{s}_\pi]_e \qquad (e=2), \tag{18.5}$$

$$F^e_{v\pi,sd} = [s^\dagger_v \otimes \tilde{d}_\pi]_e - (-1)^e [d^\dagger_\pi \otimes \tilde{s}_v]_e \qquad (e=2). \tag{18.6}$$

There are two types of orthogonal subalgebras. In the first type, the subalgebra is still a two-fluid Lie algebra. The generators $\{G^e_{\pi\pi,dd}, G^e_{vv,dd}, F^e_{\pi v,dd}\}$ constitute the two-fluid algebra $o^{\pi v}_d(10)$. The second type is that in which the subalgebra is a one-fluid Lie algebra. The generators $\{G^e_{\pi\pi,dd}, F^e_{\pi\pi,sd}\}$ and $\{G^e_{vv,dd}, F^e_{vv,sd}\}$ span the one-fluid two-levels algebras $o^\pi_{sd}(6)$, and $o^v_{sd}(6)$, respectively. The STs from $o^\pi_{sd}(6)$ and $o^v_{sd}(6)$ commute and form the $o^\pi_{sd}(6) \otimes o^v_{sd}(6)$ algebra. This survey implies the orthogonal subalgebra chains in figure 1. We have the following schemes

$$o^\pi_d(5) \subset o^\pi_{sd}(6), \qquad o^v_d(5) \subset o^v_{sd}(6), \qquad o^\pi_{sd}(6) \otimes o^v_{sd}(6) \subset o^{\pi v}_{sd}(12), \tag{19.1}$$

$$o^\pi_d(5) \otimes o^v_d(5) \subset o^{\pi v}_d(10), \qquad o^{\pi v}_s(2) \otimes o^{\pi v}_d(10) \subset o^{\pi v}_{sd}(12). \tag{19.2}$$

This implies the following subalgebra chains

$$o^{\pi v}_{sd}(12) \supset \left\{ \begin{array}{c} o^{\pi v}_s(2) \otimes o^{\pi v}_d(10) \\ o^\pi_{sd}(6) \otimes o^v_{sd}(6) \end{array} \right\} \supset o^\pi_d(5) \otimes o^v_d(5). \tag{20}$$

To write the quadratic CO of $o^{\pi v}_{sd}(12)$ in terms of multipole operators, the following operators are introduced

$$\bar{Q}^{\pi v}_{ss} = F^e_{\pi v,ss} \circ F^e_{\pi v,ss} = -\sum_e \Delta_e [F^e_{\pi v,ss} \otimes F^e_{\pi v,ss}]_0, \tag{21.1}$$

$$\bar{Q}^{\pi v}_{dd} = F^e_{\pi v,dd} \circ F^e_{\pi v,dd} = -\sum_e \Delta_e [F^e_{\pi v,dd} \otimes F^e_{\pi v,dd}]_0, \tag{21.2}$$

$$\bar{Q}^{\pi\pi}_{sd} = F^e_{\pi\pi,sd} \circ F^e_{\pi\pi,sd} = -\sum_e \Delta_e [F^e_{\pi\pi,sd} \otimes F^e_{\pi\pi,sd}]_0, \tag{21.3}$$

$$\bar{Q}^{vv}_{sd} = F^e_{vv,sd} \circ F^e_{vv,sd} = -\sum_e \Delta_e [F^e_{vv,sd} \otimes F^e_{vv,sd}]_0, \tag{21.4}$$

$$\bar{Q}^{\pi v}_{sd} = F^e_{\pi v,sd} \circ F^e_{\pi v,sd} = -\sum_e \Delta_e [F^e_{\pi v,sd} \otimes F^e_{\pi v,sd}]_0, \tag{21.5}$$

$$\bar{Q}^{v\pi}_{sd} = F^e_{v\pi,sd} \circ F^e_{v\pi,sd} = -\sum_e \Delta_e [F^e_{v\pi,sd} \otimes F^e_{v\pi,sd}]_0. \tag{21.6}$$

The operators (21) are the generalization of the multipole operator of the one-fluid IBM-1 to the two-fluid IBM-2. We can call these operators "orthogonal multipole operators". There are three types of multipole operators. The operators $\bar{Q}^{\pi\pi}_{sd}$ and $\bar{Q}^{vv}_{sd}$ represent the pure proton and neutron two-level multipole operators, respectively. However, the operators $\bar{Q}^{\pi v}_{sd}$ and $\bar{Q}^{v\pi}_{sd}$ are the mixed proton-neutron two-level multipole operators. The operators $\bar{Q}^{\pi v}_{ss}$ and $\bar{Q}^{\pi v}_{dd}$ are the mixed proton-neutron single-level multipole operators.

It is convenient to express the quadratic CO of $o^{\pi v}_{sd}(12)$, in terms of the COs of subalgebras in (20) and multipole operators. A brief calculation shows that

$$C_2[o_{sd}^{\pi\nu}(12)] = C_2[o_d^{\pi\nu}(10)] + C_2[o_s^{\pi\nu}(2)] + 2\{\bar{Q}_{sd}^{\pi\pi} + \bar{Q}_{sd}^{\nu\nu} + \bar{Q}_{sd}^{\pi\nu} + \bar{Q}_{sd}^{\nu\pi}\}, \tag{22.1}$$

$$C_2[o_{sd}^{\pi\nu}(12)] = C_2[o_{sd}^{\pi}(6)] + C_2[o_{sd}^{\nu}(6)] + 2\{\bar{Q}_{ss}^{\pi\nu} + \bar{Q}_{dd}^{\pi\nu} + \bar{Q}_{sd}^{\pi\nu} + \bar{Q}_{sd}^{\nu\pi}\}. \tag{22.2}$$

Table 1. The possible values of the coefficients $\sigma$ of the different isomorphic orthogonal algebra $o_{sd}^{\pi\nu}(12)$.

| $\sigma_{ss}^{\pi\nu}$ | $\sigma_{sd}^{\nu\pi}$ | $\sigma_{sd}^{\pi\pi}$ | $\sigma_{sd}^{\nu\nu}$ | $\sigma_{sd}^{\pi\nu}$ | $\sigma_{dd}^{\pi\nu}$ |
|---|---|---|---|---|---|
| − | − | − | − | − | − |
| − | − | − | + | + | + |
| − | + | + | + | + | − |
| − | + | + | − | − | + |
| + | − | + | − | + | − |
| + | − | + | + | − | + |
| + | + | − | + | − | − |
| + | + | − | − | + | + |

Given the Hamiltonian written in terms of the COs of one of the chains of algebraic lattice in figure 1, we can give the corresponding energy eigenvalue expression.

Before ending this section, it is important to refer to that Kota [12] introduced the orthogonal algebra $o_{sd}^{\pi\nu}(12)$ in the framwork of IBM-2 using F-spin representation. The odd rank generators [12] consist of the tensor product of the proton creation and neutron annihilation operators. In this case, the $o_{sd}^{\pi\nu}(12)$ can not be able to contain the subalgebra $o_{sd}^{\pi}(6) \otimes o_{sd}^{\nu}(6)$. The $o_{sd}^{\pi\nu}(12)$ and $o_d^{\pi\nu}(10)$ contain the subalgebras $o(6) \otimes o^F(2)$ and $o(5) \otimes o^F(2)$, respectively, where $F$ refers to F-spin represnation. However, in definitions (15) and (18), the odd rank generators consist of the tensor product of the proton (neutron) creation and proton (neutron) annihilation operators. So, the $o_{sd}^{\pi\nu}(12)$ and $o_d^{\pi\nu}(10)$ contain the subalgebras $o_{sd}^{\pi}(6) \otimes o_{sd}^{\nu}(6)$ and $o_d^{\pi}(5) \otimes o_d^{\nu}(5)$, respectively. Hence, we have completely different new DSs of the orthogonal algebra $o_{sd}^{\pi\nu}(12)$. The generators and DSs, (15) and (20), are very close to the original algebraic structure of IBM-2 with features of the subalgebra $o_d^{\pi\nu}(10)$.

## 3. Proton-Neutron Quasi-Spin Algebra

Duality is defined as an isomorphism between different theoretical formulations (perhaps quite distinct-looking). In theoretical physics, duality (complementarity) relationships are used to extend insights from one description of a phenomenon to a simpler dual description. If a duality relation can be developed between a comparatively unknown portion of algebra (symmetry) and some well-studied algebra, where several theorems and laws have already been established, and many techniques for finding solutions already exist, then this relation can be used to map the entire problem to "solid ground" where the problem is simpler to understand and more straightforward to solve. As a result, dualities are effective theory-building tools.

The $\mathfrak{su}(1,1)$-QSA is rank one Lie algebra with three generators. In nuclear physics, the duality of the IRs of $o(n)$ and $\mathfrak{su}(1,1)$ (the IRs can be correlated through their COs) leads to several beneficial relationships. For example, pairing force, and its associated seniority quantum number continue to play an essential role in nuclear theory. Exact algebraic solutions for the description of the QPTs in different versions of IBM by utilizing the Bethe ansatz and depending on dual algebraic structure were introduced [19-25].

In this section, the construction of proton-neutron QSA, $\mathfrak{su}_{sd}^{\pi\nu}(1,1)$, using the proton-neutron bosonic levels $s$, and $d$ will be demonstrated. The complementarity relationship of the COs of the $\mathfrak{su}_{sd}^{\pi\nu}(1,1)$ and $o_{sd}^{\pi\nu}(12)$ will be derived. Different coupling schemes to generate QSAs will be discussed. These schemes play an essential role in the exact algebraic solutions of transition Hamiltonians of the proton-neutron sd bosonic systems.

We will start by discussing the one-fluid single-level QSAs $\mathfrak{su}_i^\rho(1,1)$ ($i = s$, or $d$, $\rho = \pi$, or $\nu$). The pair creation operator ($S_{i;+}^\rho$), the pair annihilation operator ($S_{i;-}^\rho$) and the $S_{i;0}^\rho$ operator are defined as,

$$S_{i;+}^\rho = \frac{1}{2}\Delta_i[I_\rho^\dagger \otimes I_\rho^\dagger]_0, \quad S_{i;-}^\rho = \frac{1}{2}\Delta_i[\tilde{I}_\rho \otimes \tilde{I}_\rho]_0, \quad S_{i;0}^\rho = \frac{1}{4}\Delta_i\{[I_\rho^\dagger \otimes \tilde{I}_\rho]_0 + [\tilde{I}_\rho \otimes I_\rho^\dagger]_0\}, \tag{23}$$

where $I_\rho^\dagger \equiv d_\rho^\dagger$ and $\tilde{I}_\rho \equiv \tilde{d}_\rho$ if $i = d$ etc. It can be proven that the generators (23) satisfy the following commutation relations of the algebras $\mathfrak{su}_i^\rho(1,1)$:

$$[S_{i;0}^\rho, S_{j;\pm}^{\dot\rho}] = \pm\delta_{\rho,\dot\rho}\delta_{ij}S_{j;\pm}^{\dot\rho}, \quad [S_{i;+}^\rho, S_{j;-}^{\dot\rho}] = -2\delta_{\rho,\dot\rho}\delta_{ij}S_{j;0}^{\dot\rho}, \tag{24}$$

where the operators are constrained by the conditions; $(S_{i;+}^\rho)^\dagger = S_{i;-}^\rho$, $(S_{i;0}^\rho)^\dagger = S_{i;0}^\rho$.

Now, let us introduce a proton-neutron sd-QSA, $\mathfrak{su}_{sd}^{\pi\nu}(1,1)$, that is generated by the operators

$$S_{sd;+}^{\pi\nu} = \omega_s^\pi S_{s;+}^\pi + \omega_s^\nu S_{s;+}^\nu + \omega_d^\pi S_{d;+}^\pi + \omega_d^\nu S_{d;+}^\nu, \tag{25.1}$$

$$S_{sd;-}^{\pi\nu} = \omega_s^\pi S_{s;-}^\pi + \omega_s^\nu S_{s;-}^\nu + \omega_d^\pi S_{d;-}^\pi + \omega_d^\nu S_{d;-}^\nu, \tag{25.2}$$

$$S_{sd;0}^{\pi\nu} = S_{s;0}^\pi + S_{s;0}^\nu + S_{d;0}^\pi + S_{d;0}^\nu, \tag{25.3}$$

where, $\omega_s^\pi$, $\omega_s^\nu$, $\omega_d^\pi$, and $\omega_d^\nu$ obey the conditions

$$(\omega_s^\pi)^2 = (\omega_s^\nu)^2 = (\omega_d^\pi)^2 = (\omega_d^\nu)^2 = 1. \tag{26}$$

One can also show that the operators (25) satisfy the commutation relations of the QSA, $\mathfrak{su}_{sd}^{\pi\nu}(1,1)$,

$$[S_{sd;0}^{\pi\nu}, S_{sd;+}^{\pi\nu}] = S_{sd;+}^{\pi\nu}, \qquad [S_{sd;0}^{\pi\nu}, S_{sd;-}^{\pi\nu}] = -S_{sd;-}^{\pi\nu}, \qquad [S_{sd;+}^{\pi\nu}, S_{sd;-}^{\pi\nu}] = -2S_{sd;0}^{\pi\nu}. \tag{27}$$

The CO of $\mathfrak{su}_{sd}^{\pi\nu}(1,1)$ can be expressed as

$$C[\mathfrak{su}_{sd}^{\pi\nu}(1,1)] = S_{sd;0}^{\pi\nu}(S_{sd;0}^{\pi\nu} - 1) - S_{sd;+}^{\pi\nu} S_{sd;-}^{\pi\nu}, \tag{28}$$

and the operator $S_{sd;0}^{\pi\nu}$ is given by

$$S_{sd;0}^{\pi\nu} = \frac{1}{2}(N_{sd}^{\pi\nu} + \Omega_{sd}^{\pi\nu}), \qquad \Omega_{sd}^{\pi\nu} = \Omega_{sd}^\pi + \Omega_{sd}^\nu, \qquad \Omega_{sd}^\rho = \frac{n_{sd}}{2}, \qquad n_{sd} = \sum_{l=0,2} 2l+1, \qquad \rho = \pi, \nu. \tag{29}$$

Let $|k_{sd}^{\pi\nu} \mu_{sd}^{\pi\nu}\rangle$ denote basis vectors of the IR $k_{sd}^{\pi\nu}$ of $\mathfrak{su}_{sd}^{\pi\nu}(1,1)$, where $k_{sd}^{\pi\nu} = \frac{1}{4}, \frac{3}{4}, \frac{5}{4}, \ldots$ and $\mu_{sd}^{\pi\nu} = k_{sd}^{\pi\nu}, k_{sd}^{\pi\nu} + 1, \ldots$. We have,

$$C[\mathfrak{su}_{sd}^{\pi\nu}(1,1)]|k_{sd}^{\pi\nu} \mu_{sd}^{\pi\nu}\rangle = k_{sd}^{\pi\nu}(k_{sd}^{\pi\nu} - 1)|k_{sd}^{\pi\nu} \mu_{sd}^{\pi\nu}\rangle, \qquad S_{sd;0}^{\pi\nu}|k_{sd}^{\pi\nu} \mu_{sd}^{\pi\nu}\rangle = \mu_{sd}^{\pi\nu}|k_{sd}^{\pi\nu} \mu_{sd}^{\pi\nu}\rangle. \tag{30}$$

Once again, in a manner comparable to the $\mathfrak{u}^{\pi\nu}(12)$ and the $\mathfrak{o}_{sd}^{\pi\nu}(12)$, there are several coupling methods to create quasi-spin algebras from the one-fluid single-level operators $S_{i;0}^\rho$ and $S_{i;\pm}^\rho$. We devote special attention to the following two schemes

$$\mathfrak{su}_s^\pi(1,1) \otimes \mathfrak{su}_d^\pi(1,1) \supset \mathfrak{su}_{sd}^\pi(1,1) \quad \mathfrak{su}_s^\nu(1,1) \otimes \mathfrak{su}_d^\nu(1,1) \supset \mathfrak{su}_{sd}^\nu(1,1) \quad \mathfrak{su}_{sd}^\pi(1,1) \otimes \mathfrak{su}_{sd}^\nu(1,1) \supset \mathfrak{su}_{sd}^{\pi\nu}(1,1), \tag{31.1}$$

$$\mathfrak{su}_s^\pi(1,1) \otimes \mathfrak{su}_s^\nu(1,1) \supset \mathfrak{su}_s^{\pi\nu}(1,1) \quad \mathfrak{su}_d^\pi(1,1) \otimes \mathfrak{su}_d^\nu(1,1) \supset \mathfrak{su}_d^{\pi\nu}(1,1) \quad \mathfrak{su}_s^{\pi\nu}(1,1) \otimes \mathfrak{su}_d^{\pi\nu}(1,1) \supset \mathfrak{su}_{sd}^{\pi\nu}(1,1). \tag{31.2}$$

First, let us consider the subalgebra chain (31.1). We use the quasi-spin operators $\{S_{s;\pm}^\pi, S_{d;\pm}^\pi, S_{s;0}^\pi, S_{d;0}^\pi\}$ and $\{S_{s;\pm}^\nu, S_{d;\pm}^\nu, S_{s;0}^\nu, S_{d;0}^\nu\}$ to generate the one-fluid QSAs $\mathfrak{su}_{sd}^\pi(1,1)$, and $\mathfrak{su}_{sd}^\nu(1,1)$, respectively. These algebras are spanned by the operators

$$S_{sd;\pm}^\pi = \omega_s^\pi S_{s;\pm}^\pi + \omega_d^\pi S_{d;\pm}^\pi, \qquad S_{sd;0}^\pi = S_{s;0}^\pi + S_{d;0}^\pi, \tag{32.1}$$

$$S_{sd;\pm}^\nu = \omega_s^\nu S_{s;\pm}^\nu + \omega_d^\nu S_{d;\pm}^\nu, \qquad S_{sd;0}^\nu = S_{s;0}^\nu + S_{d;0}^\nu, \tag{32.2}$$

where $(\omega_s^\pi)^2 = (\omega_d^\pi)^2 = (\omega_s^\nu)^2 = (\omega_d^\nu)^2 = 1$. The CO of the $\mathfrak{su}_{sd}^\rho(1,1)$ is given by

$$C[\mathfrak{su}_{sd}^\rho(1,1)] = S_{sd;0}^\rho(S_{sd;0}^\rho - 1) - S_{sd;+}^\rho S_{sd;-}^\rho. \tag{33}$$

Furthermore, the operator $S_{sd;0}^\rho$ is related to the number operator by the relation

$$S_{sd;0}^\rho = \frac{1}{2}(N_{sd}^\rho + \Omega_{sd}^\rho), \qquad \Omega_{sd}^\rho = \frac{n_{sd}}{2}. \tag{34}$$

On the other hand, in the second chain, (31.2), we use the operators $\{S_{s;\pm}^\pi, S_{s;\pm}^\nu, S_{s;0}^\pi, S_{s;0}^\nu\}$ and $\{S_{d;\pm}^\pi, S_{d;\pm}^\nu, S_{d;0}^\pi, S_{d;0}^\nu\}$ to create the proton-neutron single-level QSAs $\mathfrak{su}_s^{\pi\nu}(1,1)$ and $\mathfrak{su}_d^{\pi\nu}(1,1)$, respectively. These algebras are generated by the operators

$$S_{s;\pm}^{\pi\nu} = \omega_s^\pi S_{s;\pm}^\pi + \omega_s^\nu S_{s;\pm}^\nu, \qquad S_{s;0}^{\pi\nu} = S_{s;0}^\pi + S_{s;0}^\nu, \tag{35.1}$$

$$S_{d;\pm}^{\pi\nu} = \omega_d^\pi S_{d;\pm}^\pi + \omega_d^\nu S_{d;\pm}^\nu, \qquad S_{d;0}^{\pi\nu} = S_{d;0}^\pi + S_{d;0}^\nu, \tag{35.2}$$

where $(\omega_s^\pi)^2 = (\omega_s^\nu)^2 = (\omega_d^\pi)^2 = (\omega_d^\nu)^2 = 1$. The COs of $\mathfrak{su}_i^{\pi\nu}(1,1)$, $i = s$ or $d$ are

$$C[\mathfrak{su}_i^{\pi\nu}(1,1)] = S_{i;0}^{\pi\nu}(S_{i;0}^{\pi\nu} - 1) - S_{i;+}^{\pi\nu} S_{i;-}^{\pi\nu}. \tag{36}$$

The operator $S_{i;0}^{\pi\nu}$ can be written as

$$S_{i;0}^{\pi\nu} = \frac{1}{2}(N_i^{\pi\nu} + \Omega_i^{\pi\nu}), \qquad \Omega_i^{\pi\nu} = \Omega_i^\pi + \Omega_i^\nu, \qquad \Omega_i^\rho = \frac{n_i}{2}, \qquad \rho = \pi, \nu. \tag{37}$$

Now, we can derive the duality relation between the IRs of the $\mathfrak{o}_{sd}^{\pi\nu}(12)$ and the proton-neutron QSA, $\mathfrak{su}_{sd}^{\pi\nu}(1,1)$. From (25.1), and (25.2), the product $S_{sd;+}^{\pi\nu} S_{sd;-}^{\pi\nu}$ can be written as

$$\begin{aligned} 4S_{sd;+}^{\pi\nu} S_{sd;-}^{\pi\nu} = &\, 4(S_{s;+}^\pi S_{s;-}^\pi + S_{s;+}^\nu S_{s;-}^\nu + S_{d;+}^\pi S_{d;-}^\pi + S_{d;+}^\nu S_{d;-}^\nu) \\ &+ 4\omega_{ss}^{\pi\nu}(S_{s;+}^\pi S_{s;-}^\nu + S_{s;+}^\nu S_{s;-}^\pi) + 4\omega_{dd}^{\pi\nu}(S_{d;+}^\pi S_{d;-}^\nu + S_{d;+}^\nu S_{d;-}^\pi) \\ &+ 4\omega_{sd}^{\pi\pi}(S_{s;+}^\pi S_{d;-}^\pi + S_{d;+}^\pi S_{s;-}^\pi) + 4\omega_{sd}^{\nu\nu}(S_{s;+}^\nu S_{d;-}^\nu + S_{d;+}^\nu S_{s;-}^\nu) \\ &+ 4\omega_{sd}^{\pi\nu}(S_{s;+}^\pi S_{d;-}^\nu + S_{d;+}^\nu S_{s;-}^\pi) + 4\omega_{sd}^{\nu\pi}(S_{s;+}^\nu S_{d;-}^\pi + S_{d;+}^\pi S_{s;-}^\nu), \end{aligned} \tag{38}$$

with the following definitions: $\omega_s^\pi \omega_s^\nu = \omega_{ss}^{\pi\nu}$, $\omega_s^\pi \omega_d^\pi = \omega_{sd}^{\pi\pi}$, $\omega_s^\pi \omega_d^\nu = \omega_{sd}^{\pi\nu}$, $\omega_d^\pi \omega_d^\nu = \omega_{dd}^{\pi\nu}$, $\omega_s^\nu \omega_d^\nu = \omega_{sd}^{\nu\nu}$, $\omega_d^\pi \omega_s^\nu = \omega_{sd}^{\nu\pi}$. The product $S_{sd;+}^{\pi\nu} S_{sd;-}^{\pi\nu}$ consists of single-level terms $(S_{i;+}^\rho S_{i;-}^\rho)$ and mixed terms $(S_{i;+}^\rho S_{j;-}^\rho)$. So, let us define the new set of the operators

$$\bar{\bar{Q}}_{ss}^{\pi\nu} = 4(S_{s;+}^\pi S_{s;-}^\nu + S_{s;+}^\nu S_{s;-}^\pi), \tag{39.1}$$

$$\bar{\bar{Q}}_{dd}^{\pi\nu} = 4(S_{d;+}^\pi S_{d;-}^\nu + S_{d;+}^\nu S_{d;-}^\pi), \tag{39.2}$$

$$\bar{\bar{Q}}_{sd}^{\pi\pi} = 4(S_{s;+}^\pi S_{d;-}^\pi + S_{d;+}^\pi S_{s;-}^\pi), \tag{39.3}$$

$$\bar{\bar{Q}}_{sd}^{\nu\nu} = 4(S_{s;+}^\nu S_{d;-}^\nu + S_{d;+}^\nu S_{s;-}^\nu), \tag{39.4}$$

$$\bar{\bar{Q}}_{sd}^{\pi\nu} = 4(S_{s;+}^\pi S_{d;-}^\nu + S_{d;+}^\nu S_{s;-}^\pi), \tag{39.5}$$

$$\bar{\bar{Q}}_{sd}^{\nu\pi} = 4(S_{s;+}^\nu S_{d;-}^\pi + S_{d;+}^\pi S_{s;-}^\nu). \tag{39.6}$$

The operators $\bar{\bar{Q}}$ are related to the proton-neutron multipole operators, $\bar{Q}$, of the $\mathfrak{o}_{sd}^{\pi\nu}(12)$. So, we can call these operators "quasi-spin multipole operators". From (10), (21), and (39), the operators $\bar{\bar{Q}}$ are given by

$$\bar{\bar{Q}}_{ss}^{\pi\nu} = Q_{ss}^{\pi\nu} - \bar{Q}_{ss}^{\pi\nu}, \tag{40.1}$$

$$\bar{\bar{Q}}_{dd}^{\pi\nu} = Q_{dd}^{\pi\nu} - \bar{Q}_{dd}^{\pi\nu}, \tag{40.2}$$

$$\bar{\bar{Q}}_{sd}^{\pi\pi} = Q_{sd}^{\pi\pi} - \bar{Q}_{sd}^{\pi\pi}, \tag{40.3}$$
$$\bar{\bar{Q}}_{sd}^{\nu\nu} = Q_{sd}^{\nu\nu} - \bar{Q}_{sd}^{\nu\nu}, \tag{40.4}$$
$$\bar{\bar{Q}}_{sd}^{\pi\nu} = Q_{sd}^{\pi\nu} - \bar{Q}_{sd}^{\pi\nu}, \tag{40.5}$$
$$\bar{\bar{Q}}_{sd}^{\nu\pi} = Q_{sd}^{\nu\pi} - \bar{Q}_{sd}^{\nu\pi}. \tag{40.6}$$

Moreover, the relations between one-fluid single-level quasi-spin operators and COs for the one-fluid single-level algebras $\mathfrak{u}_i^\rho(n_i)$ and $\mathfrak{o}_i^\rho(n_i)$ are given by [3]

$$4S_{i;+}^\rho S_{i;-}^\rho = -N_i^\rho + C_2[\mathfrak{u}_i^\rho(n_i)] - \frac{1}{2}C_2[\mathfrak{o}_i^\rho(n_i)]. \tag{41}$$

The single-level terms and mixed terms of the product $S_{sd;+}^{\pi\nu}S_{sd;-}^{\pi\nu}$ can be combined to produce an expression including the COs of the algebras $\mathfrak{o}_{sd}^{\pi\nu}(12)$ and $\mathfrak{u}_{sd}^{\pi\nu}(12)$, if and only if the constants $\omega$ satisfy the following conditions:

$$\omega_{ss}^{\pi\nu} = \omega_{dd}^{\pi\nu} = \omega_{sd}^{\pi\pi} = \omega_{sd}^{\nu\nu} = \omega_{sd}^{\pi\nu} = \omega_{sd}^{\nu\pi} = 1. \tag{42}$$

From (38)-(41), we get the duality relationship between the COs of the $\mathfrak{su}_{sd}^{\pi\nu}(1,1)$ and $\mathfrak{o}_{sd}^{\pi\nu}(12)$:

$$4S_{sd;+}^{\pi\nu}S_{sd;-}^{\pi\nu} = -N_{sd}^{\pi\nu} + C_2[\mathfrak{u}_{sd}^{\pi\nu}(12)] - \frac{1}{2}C_2[\mathfrak{o}_{sd}^{\pi\nu}(12)]. \tag{43}$$

The general duality relation (43) contains several duality relations as special cases, by ignoring the neutron sector, the proton sector, or the level $s$. We have,

$$4S_{sd;+}^\pi S_{sd;-}^\pi = -N_{sd}^\pi + C_2[\mathfrak{u}_{sd}^\pi(6)] - \frac{1}{2}C_2[\mathfrak{o}_{sd}^\pi(6)], \tag{44.1}$$

$$4S_{sd;+}^\nu S_{sd;-}^\nu = -N_{sd}^\nu + C_2[\mathfrak{u}_{sd}^\nu(6)] - \frac{1}{2}C_2[\mathfrak{o}_{sd}^\nu(6)], \tag{44.2}$$

$$4S_{d;+}^{\pi\nu}S_{d;-}^{\pi\nu} = -N_d^{\pi\nu} + C_2[\mathfrak{u}_d^{\pi\nu}(10)] - \frac{1}{2}C_2[\mathfrak{o}_d^{\pi\nu}(10)]. \tag{44.3}$$

Using duality relation (43), the occupation-seniority quantum numbers $N_{sd}^{\pi\nu}$ and $\nu_{sd}^{\pi\nu}$ are related to the quasi-spin quantum numbers $k_{sd}^{\pi\nu}$ and $\mu_{sd}^{\pi\nu}$ by the relations

$$k_{sd}^{\pi\nu} = \frac{1}{2}(\nu_{sd}^{\pi\nu} + \Omega_{sd}^{\pi\nu}), \qquad \mu_{sd}^{\pi\nu} = \frac{1}{2}(N_{sd}^{\pi\nu} + \Omega_{sd}^{\pi\nu}). \tag{45}$$

Finally, by using the complementarity relationships, the correspondence may be established between the DSs and IRs of orthogonal, unitary, and quasi-spin algebras. In the $\mathfrak{o}_{sd}^{\pi\nu}(12) \leftrightarrow \mathfrak{u}_s^{\pi\nu}(2) \otimes \mathfrak{u}_d^{\pi\nu}(10)$ transitional region, we have

$$\begin{array}{c} \mathfrak{u}_{sd}^{\pi\nu}(12) \\ N_{sd}^{\pi\nu} \end{array} \supset \left\{ \begin{array}{c} \mathfrak{o}_{sd}^{\pi\nu}(12) \\ \nu_{sd}^{\pi\nu} \\ \mathfrak{u}_s^{\pi\nu}(2) \otimes \mathfrak{u}_d^{\pi\nu}(10) \\ N_s^{\pi\nu} \qquad N_d^{\pi\nu} \end{array} \right\} \supset \begin{array}{c} \mathfrak{o}_s^{\pi\nu}(2) \otimes \mathfrak{o}_d^{\pi\nu}(10) \\ \nu_s^{\pi\nu} \qquad \nu_d^{\pi\nu} \end{array} \supset \begin{array}{c} \mathfrak{o}_d^\pi(5) \otimes \mathfrak{o}_d^\nu(5) \\ \nu_d^\pi \qquad \nu_d^\nu \end{array} \supset \begin{array}{c} \mathfrak{o}_d^\pi(3) \otimes \mathfrak{o}_d^\nu(3) \\ l_d^\pi \qquad l_d^\nu \end{array} \supset \begin{array}{c} \mathfrak{o}(3) \\ l \end{array}, \tag{46.1}$$

$$\begin{array}{c} \mathfrak{su}_s^\pi(1,1) \otimes \mathfrak{su}_s^\nu(1,1) \otimes \mathfrak{su}_d^\pi(1,1) \otimes \mathfrak{su}_d^\nu(1,1) \\ k_s^\pi \qquad k_s^\nu \qquad k_d^\pi \qquad k_d^\nu \end{array} \supset \begin{array}{c} \mathfrak{su}_s^{\pi\nu}(1,1) \otimes \mathfrak{su}_d^{\pi\nu}(1,1) \\ k_s^{\pi\nu} \qquad k_d^{\pi\nu} \end{array} \supset \left\{ \begin{array}{c} \mathfrak{su}_{sd}^{\pi\nu}(1,1) \\ k_{sd}^{\pi\nu} \\ \mathfrak{u}_s^{\pi\nu}(1) \otimes \mathfrak{u}_d^{\pi\nu}(1) \\ \mu_s^{\pi\nu} \qquad \mu_d^{\pi\nu} \end{array} \right\} \supset \begin{array}{c} \mathfrak{u}_{sd}^{\pi\nu}(1) \\ \mu_{sd}^{\pi\nu} \end{array}. \tag{46.2}$$

However, in the $\mathfrak{o}_{sd}^\pi(6) \otimes \mathfrak{o}_{sd}^\nu(6) \leftrightarrow \mathfrak{u}_d^\pi(5) \otimes \mathfrak{u}_d^\nu(5)$ transitional region, we get the following correspondence

$$\begin{array}{c} \mathfrak{u}_{sd}^\pi(6) \otimes \mathfrak{u}_{sd}^\nu(6) \\ N_{sd}^\pi \qquad N_{sd}^\nu \end{array} \supset \left\{ \begin{array}{c} \mathfrak{o}_{sd}^\pi(6) \otimes \mathfrak{o}_{sd}^\nu(6) \\ \nu_{sd}^\pi \qquad \nu_{sd}^\nu \\ \mathfrak{u}_d^\pi(5) \otimes \mathfrak{u}_d^\nu(5) \\ N_d^\pi \qquad N_d^\nu \end{array} \right\} \supset \begin{array}{c} \mathfrak{o}_d^\pi(5) \otimes \mathfrak{o}_d^\nu(5) \\ \nu_d^\pi \qquad \nu_d^\nu \end{array} \supset \begin{array}{c} \mathfrak{o}_d^\pi(3) \otimes \mathfrak{o}_d^\nu(3) \\ l_d^\pi \qquad l_d^\nu \end{array} \supset \begin{array}{c} \mathfrak{o}(3) \\ l \end{array}, \tag{47.1}$$

$$\begin{array}{c} \mathfrak{su}_s^\pi(1,1) \otimes \mathfrak{su}_s^\nu(1,1) \otimes \mathfrak{su}_d^\pi(1,1) \otimes \mathfrak{su}_d^\nu(1,1) \\ k_s^\pi \qquad k_s^\nu \qquad k_d^\pi \qquad k_d^\nu \end{array} \supset \left\{ \begin{array}{c} \mathfrak{su}_{sd}^\pi(1,1) \otimes \mathfrak{su}_{sd}^\nu(1,1) \\ k_{sd}^\pi \qquad k_{sd}^\nu \\ \mathfrak{u}_d^\pi(1) \otimes \mathfrak{u}_d^\nu(1) \\ \mu_d^\pi \qquad \mu_d^\nu \end{array} \right\} \supset \begin{array}{c} \mathfrak{u}_{sd}^\pi(1) \otimes \mathfrak{u}_{sd}^\nu(1) \\ \mu_{sd}^\pi \qquad \mu_{sd}^\nu \end{array}. \tag{47.2}$$

## 4. Bethe ansatz solutions and the proton-Neutron affine Lie algebra $\widehat{\mathfrak{su}(1,1)}$

To study the QPTs in (46.1) and (46.2), let us introduce the proton-neutron infinite-dimensional algebra $\widehat{\mathfrak{su}_{sd}^{\pi\nu}(1,1)}$ that can be spanned by the operators

$$S_\pm^n = (c_s^\pi)^{2n+1}S_{s;\pm}^\pi + (c_s^\nu)^{2n+1}S_{s;\pm}^\nu + (c_d^\pi)^{2n+1}S_{d;\pm}^\pi + (c_d^\nu)^{2n+1}S_{d;\pm}^\nu, \tag{48.1}$$

$$S_0^n = (c_s^\pi)^{2n}S_{s;0}^\pi + (c_s^\nu)^{2n}S_{s;0}^\nu + (c_d^\pi)^{2n}S_{d;0}^\pi + (c_d^\nu)^{2n}S_{d;0}^\nu, \tag{48.2}$$

where the values of the parameters $c_s^\pi$, $c_s^\nu$, $c_d^\pi$ and $c_d^\nu$ are real, and $n$ takes the values $0, \pm 1, \pm 2, \ldots$. It is simple to prove that these generators fulfill the following commutation relations

$$[S_+^m, S_-^n] = -2S_0^{m+n+1}, \quad [S_0^m, S_\pm^n] = \pm S_\pm^{m+n}. \tag{49}$$

Consequently, the $\{S_\mu^m, \mu = 0, +, -; m = 0, \pm 1, \pm 2, \ldots\}$ generate an affine Lie algebra $\widetilde{\mathfrak{su}_{sd}^{\pi\nu}(1,1)}$ without central extension.

The pair annihilation operator destroys the lowest weight state $|lw\rangle$ for a given quasi-spin, so that the state $|lw\rangle$ satisfies,

$$S_{s;-}^\pi|lw\rangle = 0, \quad S_{s;-}^\nu|lw\rangle = 0, \quad S_{d;-}^\pi|lw\rangle = 0, \quad S_{d;-}^\nu|lw\rangle = 0. \tag{50}$$

Because the boson number $N$ is finite, we have

$$|lw\rangle = \Big|N_{sd}^{\pi\nu}; k_s^\pi = \frac{1}{2}\left(\nu_s^\pi + \frac{1}{2}\right), \mu_s^\pi = \frac{1}{2}\left(N_s^\pi + \frac{1}{2}\right), k_s^\nu = \frac{1}{2}\left(\nu_s^\nu + \frac{1}{2}\right), \mu_s^\nu = \frac{1}{2}\left(N_s^\nu + \frac{1}{2}\right), k_d^\pi = \frac{1}{2}\left(\nu_d^\pi + \frac{5}{2}\right), \mu_d^\pi$$
$$= \frac{1}{2}\left(N_d^\pi + \frac{5}{2}\right), k_d^\nu = \frac{1}{2}\left(\nu_d^\nu + \frac{5}{2}\right), \mu_d^\nu = \frac{1}{2}\left(N_d^\nu + \frac{5}{2}\right); n_\Delta^\pi, l_d^\pi; n_\Delta^\nu, l_d^\nu; l, M\Big\rangle, \tag{51}$$

with $N_{sd}^{\pi\nu} = \nu_s^\pi + \nu_s^\nu + \nu_d^\pi + \nu_d^\nu$, $N_d^\pi = \nu_d^\pi$, $N_d^\nu = \nu_d^\nu$, $N_s^\pi = \nu_s^\pi = 0$ or 1 and $N_s^\nu = \nu_s^\nu = 0$ or 1. It is possible to interpret $\nu$ as the number of unpaired particles since the pair annihilation operator destroys the lowest weight state for a given quasi-spin (i.e. with $k = \mu$) and this state includes $N = \nu$ particles. We also have

$$S_0^n|lw\rangle = \Lambda_0^n|lw\rangle, \tag{52}$$

with

$$\Lambda_n^0 = \frac{1}{2}(c_s^\pi)^{2n}\left(\nu_s^\pi + \frac{1}{2}\right) + \frac{1}{2}(c_s^\nu)^{2n}\left(\nu_s^\nu + \frac{1}{2}\right) + \frac{1}{2}(c_d^\pi)^{2n}\left(\nu_d^\pi + \frac{5}{2}\right) + \frac{1}{2}(c_d^\nu)^{2n}\left(\nu_d^\nu + \frac{5}{2}\right). \tag{53}$$

Using $\widetilde{\mathfrak{su}_{sd}^{\pi\nu}(1,1)}$ generators, the transition Hamiltonian between $\gamma$-unstable $\mathfrak{o}_{sd}^{\pi\nu}(12)$ DS and vibrational $\mathfrak{u}_s^{\pi\nu}(2)\otimes\mathfrak{u}_d^{\pi\nu}(10)$ DS can be written as

$$H = gS_+^0 S_-^0 + \alpha S_0^1 + \beta C_2[\mathfrak{o}_d^{\pi\nu}(10)] + \gamma_d^\pi C_2[\mathfrak{o}_d^\pi(5)] + \gamma_d^\nu C_2[\mathfrak{o}_d^\nu(5)] + \sigma_d^\pi C_2[\mathfrak{o}^\pi(3)] + \sigma_d^\nu C_2[\mathfrak{o}^\nu(3)] + \sigma C_2[\mathfrak{o}(3)], \tag{54}$$

where $g$, $\alpha$, $\beta$, $\gamma_d^\pi$, $\gamma_d^\nu$, $\sigma_d^\pi$, $\sigma_d^\nu$, and $\sigma$ are real parameters. It is simply shown that (54) is equivalent to the Hamiltonian of $\mathfrak{o}_{sd}^{\pi\nu}(12)$ DS when $c_s^\pi = c_s^\nu = c_d^\pi = c_d^\nu$, and to the Hamiltonian of $\mathfrak{u}_s^{\pi\nu}(2)\otimes\mathfrak{u}_d^{\pi\nu}(10)$ DS in the case $c_s^\pi = c_s^\nu = c_s = 0$ and $c_d^\pi = c_d^\nu = c_d \neq 0$. Hence, the $c_s \neq c_d \neq 0$ cases mainly relate to the $\mathfrak{o}_{sd}^{\pi\nu}(12) \leftrightarrow \mathfrak{u}_s^{\pi\nu}(2)\otimes\mathfrak{u}_d^{\pi\nu}(10)$ transitional region. In the following, the value of the parameter $c_d$ will be fixed and allow $c_s$ to change between 0 and $c_d$.

To find eigenvectors of (54), we can utilize the Laurent series expansion of the eigenfunctions of (54) in terms of the spectral parameter $x_i$ ($i = 1, 2, \ldots, k$), using the $\widetilde{\mathfrak{su}_{sd}^{\pi\nu}(1,1)}$ generators. So, the eigenvectors of the Hamiltonian (54) can be simply written as,

$$|k; \nu_{sd}^{\pi\nu}; \nu_s^{\pi\nu}\nu_d^{\pi\nu}; \nu_s^\pi \nu_s^\nu \nu_d^\pi \nu_d^\nu; n_\Delta^\pi, l_d^\pi; n_\Delta^\nu, l_d^\nu; l, M\rangle = \sum_{n_i \in \mathbb{Z}} a_{n_1 n_2 \ldots n_k} x_1^{n_1} x_2^{n_2} \ldots x_k^{n_k}|lw\rangle. \tag{55}$$

Because the wavefunctions are analytic functions, then all $a_{-1} = a_{-2} = a_{-3} = \cdots = 0$, and the Laurent series will be reduced to a Tayler series, i.e., $n$ is zero or a positive integer. Using (49), it can be shown that all coefficients $a_{n_1 n_2 \ldots n_k}$ in (55) are equal to 1. Thus, the wavefunctions (55) can be represented as

$$|k; \nu_{sd}^{\pi\nu}; \nu_s^{\pi\nu}\nu_d^{\pi\nu}; \nu_s^\pi \nu_s^\nu \nu_d^\pi \nu_d^\nu; n_\Delta^\pi, l_d^\pi; n_\Delta^\nu, l_d^\nu; l, M\rangle = \mathfrak{N} S_+^{x_1} S_+^{x_2} \ldots S_+^{x_k}|lw\rangle, \tag{56}$$

where $\mathfrak{N}$ is a normalization factor and $S_+^{x_i}$ ($i = 1, 2, \ldots, k$) is a functional operator of $S_{s;+}^\pi$, $S_{s;+}^\nu$, $S_{d;+}^\pi$ and $S_{d;+}^\nu$ with spectral parameter $x_i$,

$$S_+^{x_i} = \frac{c_s^\pi}{1-(c_s^\pi)^2 x_i}S_{s;+}^\pi + \frac{c_s^\nu}{1-(c_s^\nu)^2 x_i}S_{s;+}^\nu + \frac{c_d^\pi}{1-(c_d^\pi)^2 x_i}S_{d;+}^\pi + \frac{c_d^\nu}{1-(c_d^\nu)^2 x_i}S_{d;+}^\nu. \tag{57}$$

The spectral parameters $x_i$ can be calculated by using the following set of Bethe ansatz equations,

$$\frac{\alpha}{x_i} = \frac{g(c_s^\pi)^2\left(\nu_s^\pi+\frac{1}{2}\right)}{1-(c_s^\pi)^2 x_i} + \frac{g(c_s^\nu)^2\left(\nu_s^\nu+\frac{1}{2}\right)}{1-(c_s^\nu)^2 x_i} + \frac{g(c_d^\pi)^2\left(\nu_d^\pi+\frac{5}{2}\right)}{1-(c_d^\pi)^2 x_i} + \frac{g(c_d^\nu)^2\left(\nu_d^\nu+\frac{5}{2}\right)}{1-(c_d^\nu)^2 x_i} - \sum_{j\neq i}\frac{2g}{x_i-x_j}, \text{for } i = 1,2,\ldots,k. \tag{58}$$

The energy spectra $E^{(k)}$ of the Hamiltonian (54) can then be written as

$$E^k = \sum_i \frac{\alpha}{x_i} + \beta \nu_d^{\pi\nu}(\nu_d^{\pi\nu}+8) + \gamma_d^\pi \nu_d^\pi(\nu_d^\pi+3) + \gamma_d^\nu \nu_d^\nu(\nu_d^\nu+3)$$
$$+\sigma_d^\pi l_d^\pi(l_d^\pi+1) + \sigma_d^\nu l_d^\nu(l_d^\nu+1) + \sigma l(l+1) + \alpha \Lambda_1^0, \tag{59}$$

where

$$\Lambda_1^0 = \frac{1}{2}(c_s^\pi)^2\left(\nu_s^\pi+\frac{1}{2}\right) + \frac{1}{2}(c_s^\nu)^2\left(\nu_s^\nu+\frac{1}{2}\right) + \frac{1}{2}(c_d^\pi)^2\left(\nu_d^\pi+\frac{5}{2}\right) + \frac{1}{2}(c_d^\nu)^2\left(\nu_d^\nu+\frac{5}{2}\right). \tag{60}$$

The relationship between the quantum number $k$ and the total boson number $N_{sd}^{\pi\nu}$ is given by

$$N_{sd}^{\pi\nu} = 2k + \nu_s^\pi + \nu_s^\nu + \nu_d^\pi + \nu_d^\nu. \tag{61}$$

As a result, the quantum numbers $\nu_s^\pi$, $\nu_s^\nu$ and $\nu_s^{\pi\nu}$ could be ignored for a bosonic system with a fixed number of bosons. To differentiate various solutions of (54), an extra quantum number, $\xi = 1, 2, \ldots, p$, should be added.

Table 2. Parameters of the IBM-2 Hamiltonian, (54), are extracted by least-square fitting from the experimental data for different nuclei. $\alpha = \beta, \gamma_d^\pi, \gamma_d^\nu, \sigma_d^\pi, \sigma_d^\nu, g = 1$, and $\sigma$ are the parameters of transitional Hamiltonian for each nucleus. $c_s$ and $c_d$ are the coefficients of the algebra $\mathfrak{su}_{sd}^{\pi\nu}(1,1)$.

|  | $\alpha = \beta$ | $\gamma_d^\pi$ | $\gamma_d^\nu$ | $\sigma_d^\pi$ | $\sigma_d^\nu$ | $\sigma$ | $c_s$ | $c_d$ |
|---|---|---|---|---|---|---|---|---|
| $^{70}$Ge | 54.0 | 4.00 | 19.0 | 5.00 | 22.0 | 24.0 | $10^{-6}$ | 1.0 |
| $^{76}$Se | 46.0 | 5.00 | 7.00 | 1.00 | 9.00 | 10.0 | $10^{-6}$ | 1.0 |
| $^{78}$Se | 44.0 | 42.0 | 12.0 | 6.00 | 11.0 | 11.0 | $10^{-6}$ | 1.0 |
| $^{96}$Mo | 53.0 | 10.0 | 3.00 | 2.00 | 14.0 | 10.0 | $10^{-6}$ | 1.0 |
| $^{98}$Mo | 57.0 | 6.00 | 2.00 | 2.00 | 8.00 | 6.00 | $10^{-6}$ | 1.0 |
| $^{100}$Ru | 46.0 | 45.0 | 2.00 | 1.00 | 5.00 | 5.00 | $10^{-6}$ | 1.0 |
| $^{102}$Ru | 37.0 | 2.00 | 17.0 | 1.00 | 6.00 | 8.00 | $10^{-6}$ | 1.0 |

Table 3. Comparison between experimental and calculated (IBM-2) energies (in keV) for the positive parity states of the $^{70}$Ge isotopes. The experimental data are taken from [33] for $^{70}$Ge. The seniority quantum numbers and the number of roots $k$ are listed. The values of $L_d^\pi, L_d^\nu$, and $L^+$ indicate the spins of the corresponding levels. The deviation σ monitors the quality of the fitting. All quantities are expressed in keV.

| | | | | | | $^{70}$Ge | |
|---|---|---|---|---|---|---|---|
| $k$ | $\nu_d^\pi$ | $\nu_d^\nu$ | $l_d^\pi$ | $l_d^\nu$ | $l^+$ | $\epsilon_{\text{exp}}$ | $\epsilon_{\text{th}}$ |
| 3 | 0 | 0 | 0 | 0 | $0^+$ | 0.00000 | 0.00000 |
| 3 | 0 | 1 | 0 | 2 | $2^+$ | 1039.50 | 865.000 |
| 2 | 1 | 1 | 2 | 2 | $0^+$ | 1215.60 | 1388.00 |
| 2 | 0 | 2 | 0 | 2 | $2^+$ | 1707.70 | 1600.00 |
| 2 | 1 | 1 | 2 | 2 | $4^+$ | 2153.10 | 1868.00 |
| 1 | 1 | 2 | 2 | 2 | $0^+$ | 2307.00 | 2231.00 |
| 1 | 1 | 2 | 2 | 2 | $2^+$ | 2156.70 | 2375.00 |
| 1 | 1 | 2 | 2 | 2 | $3^+$ | 2451.30 | 2519.00 |
| 1 | 1 | 2 | 2 | 4 | $2^+$ | 2534.90 | 2683.00 |
| 1 | 1 | 2 | 2 | 2 | $4^+$ | 2806.30 | 2711.00 |
| 1 | 1 | 2 | 2 | 4 | $3^+$ | 3046.40 | 2827.00 |
| 1 | 0 | 3 | 0 | 4 | $4^+$ | 3058.60 | 3125.00 |
| 1 | 1 | 2 | 2 | 4 | $4^+$ | 3194.20 | 3019.00 |
| 1 | 1 | 2 | 2 | 4 | $6^+$ | 3296.90 | 3547.00 |
| σ | | | | | | | 172.73 |

## 5. Theoretical results and comparison with experimental counterpart

It is critical to determine whether there is experimental data that supports such DSs expectations. We focus on the vibrational $\mathfrak{u}_s^{\pi\nu}(2) \otimes \mathfrak{u}_d^{\pi\nu}(10)$ DS in $^{70}$Ge, $^{76-78}$Se, $^{96-98}$Mo, and $^{100-102}$Ru isotopes. The $^{76-78}$Se isotopes have nearly spherical level structures with vibrational characteristics. The two-phonon triplet levels in $^{76-78}$Se isotopes are approximately at twice the energy of the $2^+$ state. However, deviations from a pure vibrational phonon model have been addressed, for instance, in $^{78}$Se, non-zero values of the quadrupole moment of the initial $2^+$ state were reported [28]. The vibrational limit of IBM-2 computation [29] provided appropriately the energy levels except for the fact that the level $0^+$ was very high (by 200 keV at least) compared with the experimental values. Also, the $^{96-98}$Mo isotopes have served as good examples of quadrupole vibrational nuclei, with $\mathfrak{u}(5)$ DS. Such nuclei are found near the closed-shell containing intruder and normal states which are positioned throughout the $Z = 40$ sub-shell. The description of low-lying energy levels of the Mo isotopes was given using the collective model [30]. The microscopic investigations [31] reveal that IBM-2 was the best model to describe the spectra of $^{98}$Mo, where there is a mixing between γ-unstable and vibrational DSs. The $^{100-102}$Ru isotopes are characterized by the energy ratio $E(4_1^+)/E(2_1^+) \approx 2$, states of the two-phonon triplet exist at around $2E(2_1^+)$, and the three-phonon quintuplet states are approximately at $3E(2_1^+)$. The description of the low-lying spectra in the $^{100-102}$Ru isotopes was investigated using IBM-1 and IBM-2 [32].

Eigenvalues and eigenvectors of the $o_{sd}^{\pi\nu}(12) \leftrightarrow \mathfrak{u}_s^{\pi\nu}(2) \otimes \mathfrak{u}_d^{\pi\nu}(10)$ transition Hamiltonian are determined by solving Bethe ansatz equations (58) with the fitting process to the experimental data to get the parameters of the Hamiltonian. The parameters of Hamiltonian are shown in table 2. In tables 3-6, available experimental results for these nuclei are compared to model predictions. The control parameters $c_s$ have been proposed in the interval $[0,1]$. The accuracy of the fitting for each nucleus is measured by the root mean absolute error, which is given by $\sigma = \sqrt{\sum_i |\epsilon_{\text{exp}} - \epsilon_{\text{th}}|^2 / (n-1)}$, where $\epsilon_{\text{th}}$ and $\epsilon_{\text{exp}}$ are the theoretical and experimental energy levels, and $n$ is the number of levels that are considered in the fitting process. The present results for the control parameter $c_s$ suggest that these isotopes are the good candidates for the vibrational $\mathfrak{u}_s^{\pi\nu}(2) \otimes \mathfrak{u}_d^{\pi\nu}(10)$ DS.

Table 4. The same as in table 3 but for the $^{76}$Se and $^{78}$Se isotopes. The experimental data are taken from [34] for $^{76}$Se and [35] for $^{78}$Se.

| | | | | | | $^{76}$Se | | | | | | | | $^{78}$Se | |
|---|---|---|---|---|---|---|---|---|---|---|---|---|---|---|---|
| $k$ | $v_d^\pi$ | $v_d^\nu$ | $l_d^\pi$ | $l_d^\nu$ | $l^+$ | $\epsilon_{\text{exp}}$ | $\epsilon_{\text{th}}$ | $k$ | $v_d^\pi$ | $v_d^\nu$ | $l_d^\pi$ | $l_d^\nu$ | $l^+$ | $\epsilon_{\text{exp}}$ | $\epsilon_{\text{th}}$ |
| 3 | 0 | 0 | 0 | 0 | $0^+$ | 0.00000 | 0.00000 | 2 | 0 | 0 | 0 | 0 | $0^+$ | 0.00000 | 0.00000 |
| 3 | 0 | 1 | 0 | 2 | $2^+$ | 559.100 | 578.93 | 2 | 0 | 1 | 0 | 2 | $2^+$ | 613.700 | 598.00 |
| 2 | 1 | 1 | 2 | 2 | $0^+$ | 1122.30 | 1073.86 | 2 | 1 | 1 | 2 | 2 | $0^+$ | 1498.50 | 1242.00 |
| 2 | 0 | 2 | 0 | 2 | $2^+$ | 1216.20 | 1149.86 | 1 | 0 | 2 | 0 | 2 | $2^+$ | 1308.60 | 1176.00 |
| 2 | 1 | 1 | 2 | 2 | $4^+$ | 1330.90 | 1273.86 | 1 | 1 | 1 | 2 | 2 | $4^+$ | 1502.80 | 1462.00 |
| 1 | 1 | 2 | 2 | 2 | $2^+$ | 1787.00 | 1796.80 | 1 | 1 | 2 | 2 | 2 | $3^+$ | 1853.90 | 2040.00 |
| 1 | 1 | 2 | 2 | 2 | $3^+$ | 1688.90 | 1856.80 | 1 | 1 | 2 | 2 | 4 | $2^+$ | 1995.90 | 2128.00 |
| 1 | 1 | 2 | 2 | 2 | $4^+$ | 2025.90 | 1936.80 | 1 | 0 | 3 | 0 | 4 | $4^+$ | 2190.70 | 2174.00 |
| 1 | 1 | 2 | 2 | 4 | $6^+$ | 2262.40 | 2282.80 | 1 | 1 | 2 | 2 | 4 | $6^+$ | 2546.50 | 2524.00 |
| $\sigma$ | | | | | | | 76.676 | | | | | | | | 131.413 |

Table 5. The same as in table 3 but for the $^{96}$Mo and $^{98}$Mo isotopes. The experimental data are taken from [36] for $^{96}$Mo and [37] for $^{98}$Mo.

| | | | | | | $^{96}$Mo | | | | | | | | $^{98}$Mo | |
|---|---|---|---|---|---|---|---|---|---|---|---|---|---|---|---|
| $k$ | $v_d^\pi$ | $v_d^\nu$ | $l_d^\pi$ | $l_d^\nu$ | $l^+$ | $\epsilon_{\text{exp}}$ | $\epsilon_{\text{th}}$ | $k$ | $v_d^\pi$ | $v_d^\nu$ | $l_d^\pi$ | $l_d^\nu$ | $l^+$ | $\epsilon_{\text{exp}}$ | $\epsilon_{\text{th}}$ |
| 2 | 0 | 0 | 0 | 0 | $0^+$ | 0.00000 | 0.00000 | 3 | 0 | 0 | 0 | 0 | $0^+$ | 0.00000 | 0.00000 |
| 2 | 0 | 1 | 0 | 2 | $2^+$ | 778.000 | 659.500 | 3 | 0 | 1 | 0 | 2 | $2^+$ | 787.380 | 633.500 |
| 1 | 1 | 1 | 2 | 2 | $0^+$ | 1148.00 | 1261.00 | 2 | 0 | 2 | 0 | 2 | $2^+$ | 1432.20 | 1301.00 |
| 2 | 1 | 1 | 2 | 2 | $0^+$ | 1330.00 | 1261.00 | 2 | 1 | 1 | 2 | 2 | $4^+$ | 1510.00 | 1409.00 |
| 1 | 0 | 2 | 0 | 2 | $2^+$ | 1497.00 | 1287.00 | 1 | 0 | 3 | 0 | 0 | $0^+$ | 1963.00 | 2002.50 |
| 1 | 1 | 1 | 2 | 2 | $2^+$ | 1625.00 | 1321.00 | 1 | 1 | 2 | 2 | 2 | $0^+$ | 2037.50 | 2070.50 |
| 1 | 1 | 1 | 2 | 2 | $4^+$ | 1628.00 | 1461.00 | 1 | 1 | 2 | 2 | 2 | $3^+$ | 2104.00 | 2142.50 |
| 1 | 0 | 2 | 0 | 4 | $4^+$ | 1869.00 | 1623.00 | 1 | 1 | 2 | 2 | 4 | $2^+$ | 2206.60 | 2218.50 |
| 1 | 1 | 2 | 2 | 2 | $3^+$ | 1978.00 | 2114.50 | 1 | 1 | 2 | 2 | 2 | $4^+$ | 2223.80 | 2190.50 |
| 1 | 1 | 2 | 2 | 4 | $2^+$ | 2095.00 | 2250.50 | 1 | 0 | 3 | 0 | 4 | $4^+$ | 2240.00 | 2282.50 |
| 1 | 0 | 3 | 0 | 4 | $4^+$ | 2219.00 | 2362.50 | 1 | 1 | 2 | 2 | 4 | $4^+$ | 2333.50 | 2302.50 |
| 1 | 1 | 2 | 2 | 4 | $4^+$ | 2476.00 | 2390.50 | 1 | 1 | 2 | 2 | 4 | $6^+$ | 2343.60 | 2434.50 |
| 1 | 1 | 2 | 2 | 4 | $5^+$ | 2438.00 | 2490.50 | | | | | | | | |
| 1 | 1 | 2 | 2 | 4 | $6^+$ | 2440.00 | 2610.50 | | | | | | | | |
| $\sigma$ | | | | | | | 166.187 | | | | | | | | 78.338 |

Table 6. The same as in table 3 but for the $^{100}$Ru and $^{102}$Ru isotopes. The experimental data are taken from [38] for $^{100}$Ru and [39] for $^{102}$Ru.

| | | | | | | $^{100}$Ru | | | | | | | | $^{102}$Ru | |
|---|---|---|---|---|---|---|---|---|---|---|---|---|---|---|---|
| $k$ | $v_d^\pi$ | $v_d^\nu$ | $l_d^\pi$ | $l_d^\nu$ | $l^+$ | $\epsilon_{\text{exp}}$ | $\epsilon_{\text{th}}$ | $k$ | $v_d^\pi$ | $v_d^\nu$ | $l_d^\pi$ | $l_d^\nu$ | $l^+$ | $\epsilon_{\text{exp}}$ | $\epsilon_{\text{th}}$ |
| 2 | 0 | 0 | 0 | 0 | $0^+$ | 0.00000 | 0.00000 | 3 | 0 | 0 | 0 | 0 | $0^+$ | 0.00000 | 0.00000 |
| 2 | 0 | 1 | 0 | 2 | $2^+$ | 539.500 | 505.000 | 3 | 0 | 1 | 0 | 2 | $2^+$ | 475.100 | 503.500 |
| 2 | 1 | 1 | 2 | 2 | $0^+$ | 1130.30 | 1190.00 | 2 | 1 | 1 | 2 | 2 | $0^+$ | 943.700 | 895.000 |
| 1 | 1 | 1 | 2 | 2 | $2^+$ | 1362.20 | 1220.00 | 2 | 0 | 2 | 0 | 2 | $2^+$ | 1103.10 | 1031.00 |
| 1 | 1 | 1 | 2 | 2 | $4^+$ | 1226.50 | 1290.00 | 2 | 1 | 1 | 2 | 2 | $4^+$ | 1106.40 | 1055.00 |
| 1 | 1 | 2 | 2 | 2 | $3^+$ | 1881.00 | 1883.00 | 1 | 1 | 2 | 2 | 2 | $3^+$ | 1521.60 | 1592.50 |
| 1 | 1 | 2 | 2 | 4 | $2^+$ | 1865.10 | 1923.00 | 1 | 1 | 2 | 2 | 4 | $2^+$ | 1580.60 | 1628.50 |
| 1 | 1 | 2 | 2 | 4 | $4^+$ | 2062.60 | 1993.00 | 1 | 1 | 2 | 2 | 4 | $4^+$ | 1799.00 | 1740.50 |
| 1 | 1 | 2 | 2 | 4 | $6^+$ | 2075.60 | 2103.00 | 1 | 1 | 2 | 2 | 4 | $6^+$ | 1873.30 | 1916.50 |
| $\sigma$ | | | | | | | 68.882 | | | | | | | | 54.346 |

## 6. Conclusion

In this paper, we explored the richer possibilities that the proton-neutron unitary, $\mathfrak{u}_{sd}^{\pi\nu}(12)$, orthogonal, $\mathfrak{o}_{sd}^{\pi\nu}(12)$, and quasi-spin $\mathfrak{su}_{sd}^{\pi\nu}(1,1)$ algebras of the sd bosonic system provide. There are two important types of subalgebras for the $\mathfrak{u}_{sd}^{\pi\nu}(12)$, $\mathfrak{o}_{sd}^{\pi\nu}(12)$, and $\mathfrak{su}_{sd}^{\pi\nu}(1,1)$ algebras. In the first type, the subalgebras are already one-fluid two-level algebras: $(\mathfrak{u}_{sd}^{\pi}(6), \mathfrak{o}_{sd}^{\pi}(6), \mathfrak{su}_{sd}^{\pi}(1,1))$ and $(\mathfrak{u}_{sd}^{\nu}(6), \mathfrak{o}_{sd}^{\nu}(6), \mathfrak{su}_{sd}^{\nu}(1,1))$. The second type is that in which the subalgebras are still two-fluid but one-level algebras: $(\mathfrak{u}_{d}^{\pi\nu}(10), \mathfrak{o}_{d}^{\pi\nu}(10), \mathfrak{su}_{d}^{\pi\nu}(1,1))$ and $(\mathfrak{u}_{s}^{\pi\nu}(2), \mathfrak{o}_{s}^{\pi\nu}(2),$ and $\mathfrak{su}_{s}^{\pi\nu}(1,1))$. The main difference between these two algebraic structures is the multipole operators. We introduce three types of multipole operators, namely, unitary, orthogonal, and quasi-spin multipole operators. The relationships between multipole operators of the $\mathfrak{u}_{sd}^{\pi\nu}(12)$, $\mathfrak{o}_{sd}^{\pi\nu}(12)$, and $\mathfrak{su}_{sd}^{\pi\nu}(1,1)$ algebras are given. The complementarity relationship of the COs of the $\mathfrak{su}_{sd}^{\pi\nu}(1,1)$ and $\mathfrak{o}_{sd}^{\pi\nu}(12)$ is derived. These new DSs are very close to the original algebraic structure of IBM-2 with features of the subalgebras $\mathfrak{o}_{d}^{\pi\nu}(10), \mathfrak{u}_{d}^{\pi\nu}(10), \mathfrak{su}_{d}^{\pi\nu}(1,1)$.

These relations allow us to utilize an infinite-dimensional algebraic approach to get exact solutions for energy eigenvalues and eigenvectors for nuclei in the $\mathfrak{o}_{sd}^{\pi\nu}(12) \leftrightarrow \mathfrak{u}_{s}^{\pi\nu}(2) \otimes \mathfrak{u}_{d}^{\pi\nu}(10)$ transition region within the framework of the IBM-2. Only one- and two-body interactions were considered in the Hamiltonian. The total number of bosons and angular momentum were assumed to be conserved quantities. Low-lying energy levels of the $^{70}$Ge, $^{76\text{-}78}$Se, $^{96\text{-}98}$Mo, and $^{100\text{-}102}$Ru isotopes were computed in the $\mathfrak{u}_{s}^{\pi\nu}(2) \otimes \mathfrak{u}_{d}^{\pi\nu}(10)$ DS limit. For the nuclei under consideration, the theoretical predictions are in good agreement with the available empirical results. Searching for nuclei near the $\mathfrak{o}_{sd}^{\pi\nu}(12)$ limit will be a part of future work.